\documentclass[aps,twocolumn,showpacs]{revtex4}
\usepackage{graphicx}
\usepackage{textcomp}
\usepackage{psfrag}

\begin{document}
\title{Hidden Order in Crackling Noise during Peeling of an Adhesive Tape }
\author{Jagadish Kumar$^{1}$}
\author{M. Ciccotti$^{2}$}  
\author{G. Ananthakrishna$^{1}$}
\affiliation{{$^1$ Materials Research Centre 
Indian Institute of Science Bangalore-560012, India.\\} 
$^2$ Laboratoire des Collo\"ides Verres et Nanomat\'eriaux CNRS UMR 5587 
Universit\'e de Montpellier II, Place Bataillon, 34095 Montpellier CEDEX 5, 
France\\}

\begin{abstract}
We address the long standing problem of recovering dynamical information from noisy acoustic emission signals arising from peeling of an adhesive tape subject to constant traction velocity.  Using phase space reconstruction procedure we demonstrate the deterministic chaotic dynamics by establishing the existence of correlation dimension as also a positive Lyapunov exponent in a mid range of traction velocities. The results are explained on the basis of the model that also emphasizes the deterministic origin of acoustic emission by clarifying its connection to sticks-slip dynamics.
\end{abstract}
 
\pacs{05.45.-a, 05.45.Tp, 62.20.Mk, 83.60.Df}

\maketitle
Adhesion continues to generate new directions of interest due to the wide ranging interdisciplinary  issues involved and its technological importance. For instance, the recent surge in interest can be traced to its relevance to biological systems, in particular,  the desire to design  adhesive materials that mimic fibrillar adhesion inherent to   biological species like gecko \cite{Jagota07}. Despite the progress, day-to-day experience like acoustic emission (AE) during peeling of an adhesive tape has remained ill explained.  This can be traced to the fact that most information  is obtained from quasi-static or near steady state conditions and much less attention has been paid to nonequilibrium time dependent dissipative aspects of adhesion, and related phenomenon  like friction (which is adhesion and wear) \cite{Kendall00,Urbakh04,Persson} as also AE. As kinetic and dynamical aspects involve interplay of internal relaxation time scales (determined by  molecular mechanisms) with the applied time scale,  they are important in a variety of situations that are subject  to fluctuating forces such as flexible joints,  composites, and even dynamics of cell orientation \cite{Rumi07}. 

Dynamical information can be obtained using experiments on  peeling of an adhesive tape mounted on a roller. These experiments show that peeling is jerky accompanied by a characteristic crackling noise \cite{MB,Ciccotti04}. The jerky nature is attributed to the switching of the peel  process between two stable dissipative branches  separated by an unstable  one. ( The low and high velocity branches arise from viscous dissipation and  brittle fracture respectively.) The  negative force-velocity relation is common to many stick-slip situations, for example, sliding friction \cite{Urbakh04,Persson} and the Portevin-Le Chatelier (PLC)  effect, a plastic instability observed in tensile deformation of dilute alloys \cite{GA07}, to name only two. In general, stick-slip dynamics results from  a competition among inherent time scales \cite{GA07,Anan04}, here, the viscoelastic time scale and the time scale of the pull speed.  All stick-slip processes are examples of   deterministic nonlinear dynamics.

In contrast to stick-slip nature of peeling, the origin of AE ( even in the  general context) is ill understood. Recently,  we suggested that the energy dissipated in the form of AE can be modeled in terms of  the local displacement rate \cite{Rumiprl}. A model relevant for the experimental set up  that includes such a term reproduces  major experimental features of AE as also that of  the peel front dynamics\cite{Rumiprl}.  The model also predicts spatio-temporal chaos  for a specific set of parameters.  Moreover, it is long believed that AE and stick-slip peel dynamics  are related. But, establishing such a connection requires extracting  quantitative dynamical information from the AE signals which so far has not been possible largely due to the highly noisy nature of AE signals. Here, we show that deterministic dynamics governs the AE process by  demonstrating  the existence of  chaotic dynamics using nonlinear time series analysis. The results are explained using a model that also provides insight into the connection between AE signals and stick-slip dynamics. 
\begin{figure}
\hbox{
\includegraphics[height=1.7cm,width=4.8cm]{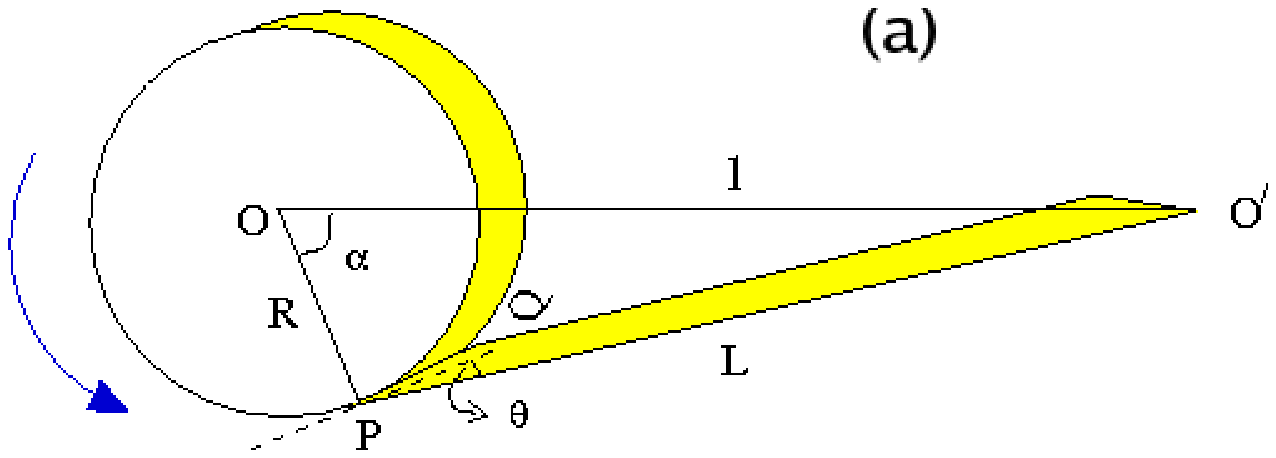}
\includegraphics[height=2.3cm,width=3.5cm]{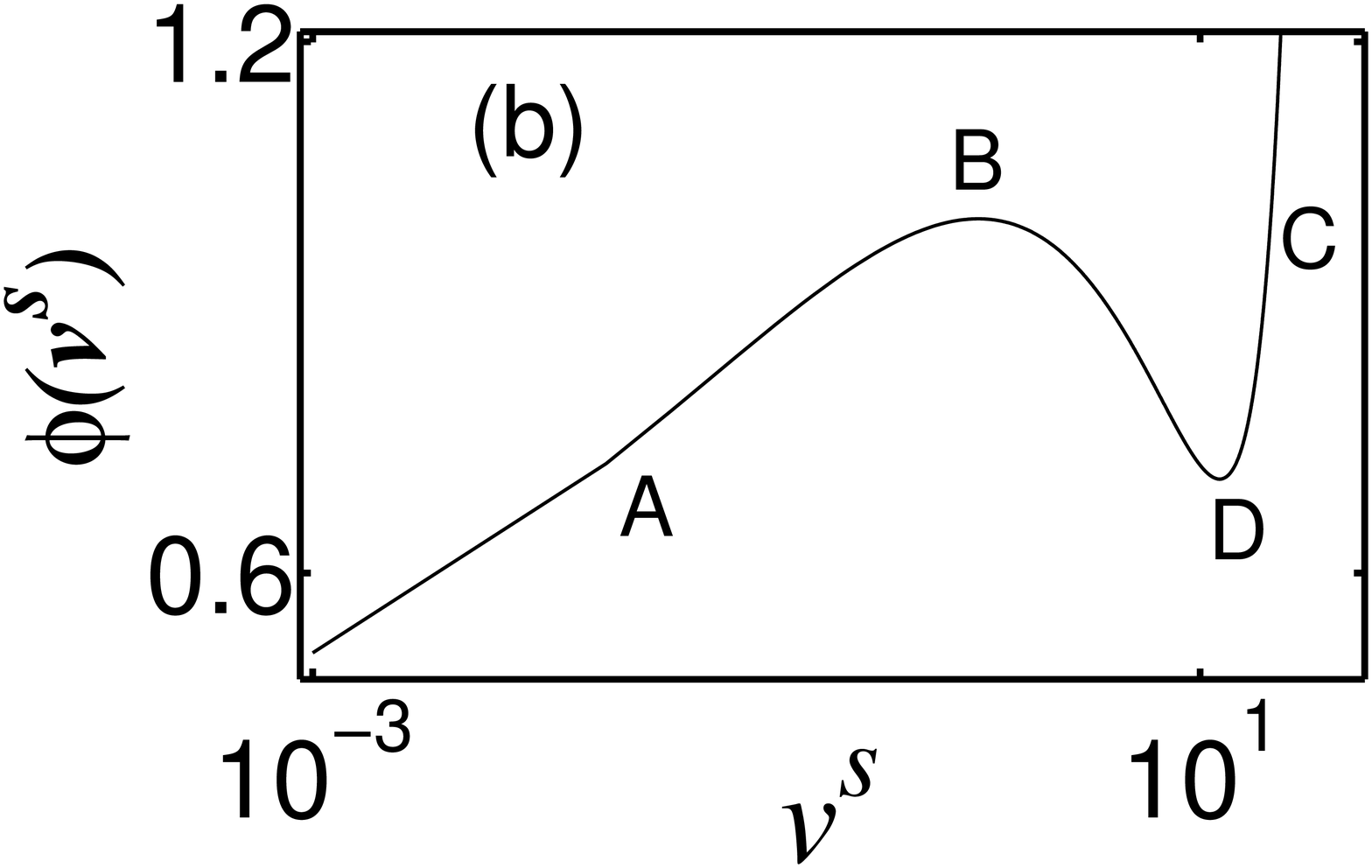}
}
\caption{(a) A schematic of the experimental setup. (b) Plot of the scaled peel force function $\phi(v^s)$ as a function of $v^s$.
}
\label{schematic}
\end{figure}

Retrieval of information about the underlying process is also important in the general context of AE as  it is observed in a large number of systems like micro-fracturing process, volcanic activity \cite{Petri94}, collective dislocation motion \cite{Miguel,Weiss} etc. However, most studies \cite{Petri94,Miguel}, except Ref. \cite{Weiss}, are simple statistical studies showing the power law distribution of AE signals as experimental realizations of self-organized criticality \cite{Bak}.  Even in Ref. \cite{Weiss}, the extracted fractal dynamics of dislocation generated AE sources is aided by use of  multiple transducers.  However, the situation is more complex in peeling experiments as only a single  transducer is used  leading to scalar AE signals that are also substantially noisy making the intended task even more challenging.

\begin{figure}
\vbox{
\hbox{
\includegraphics[height=2.5cm,width=4.1cm]{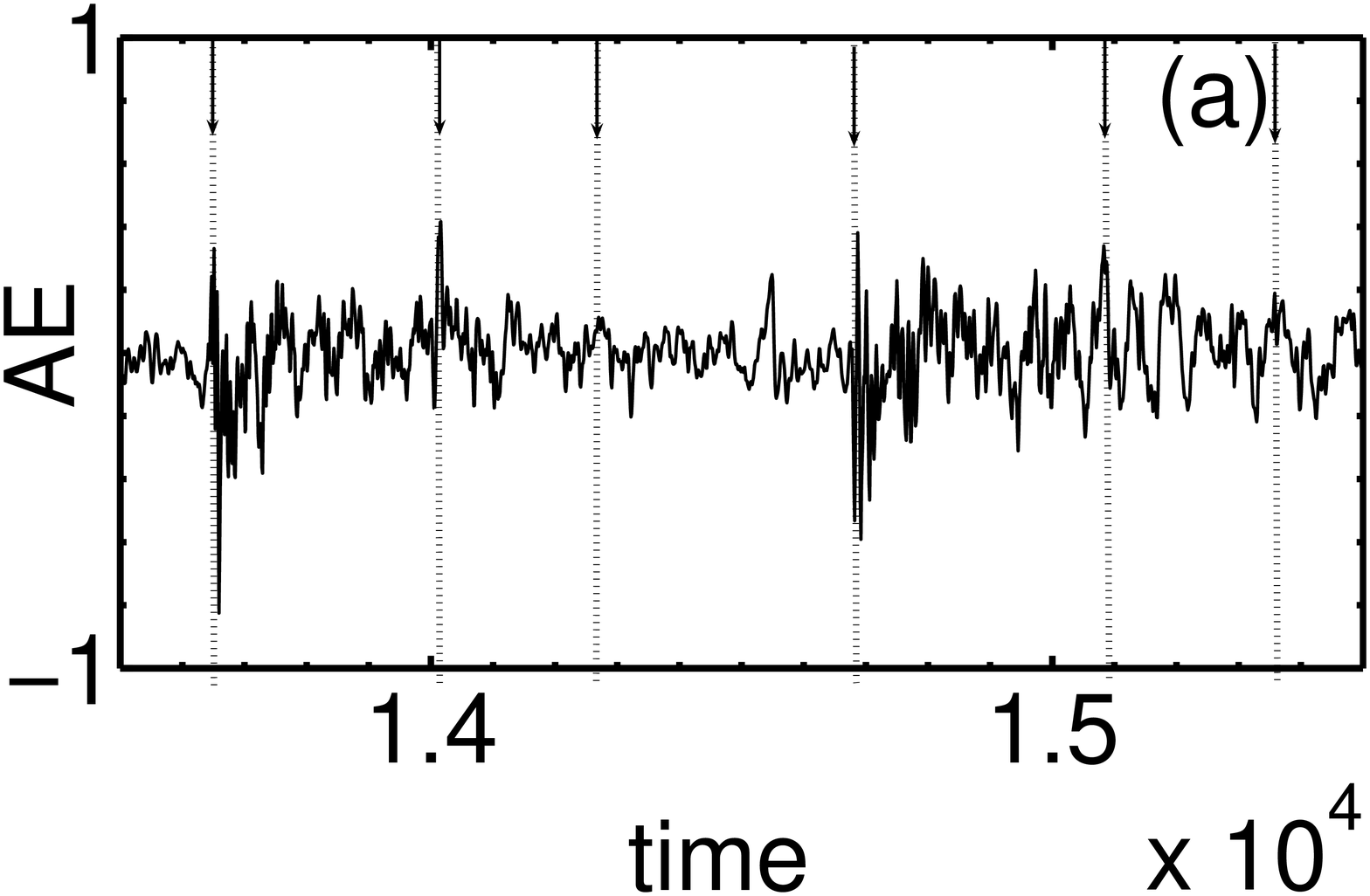}
\includegraphics[height=2.5cm,width=4.1cm]{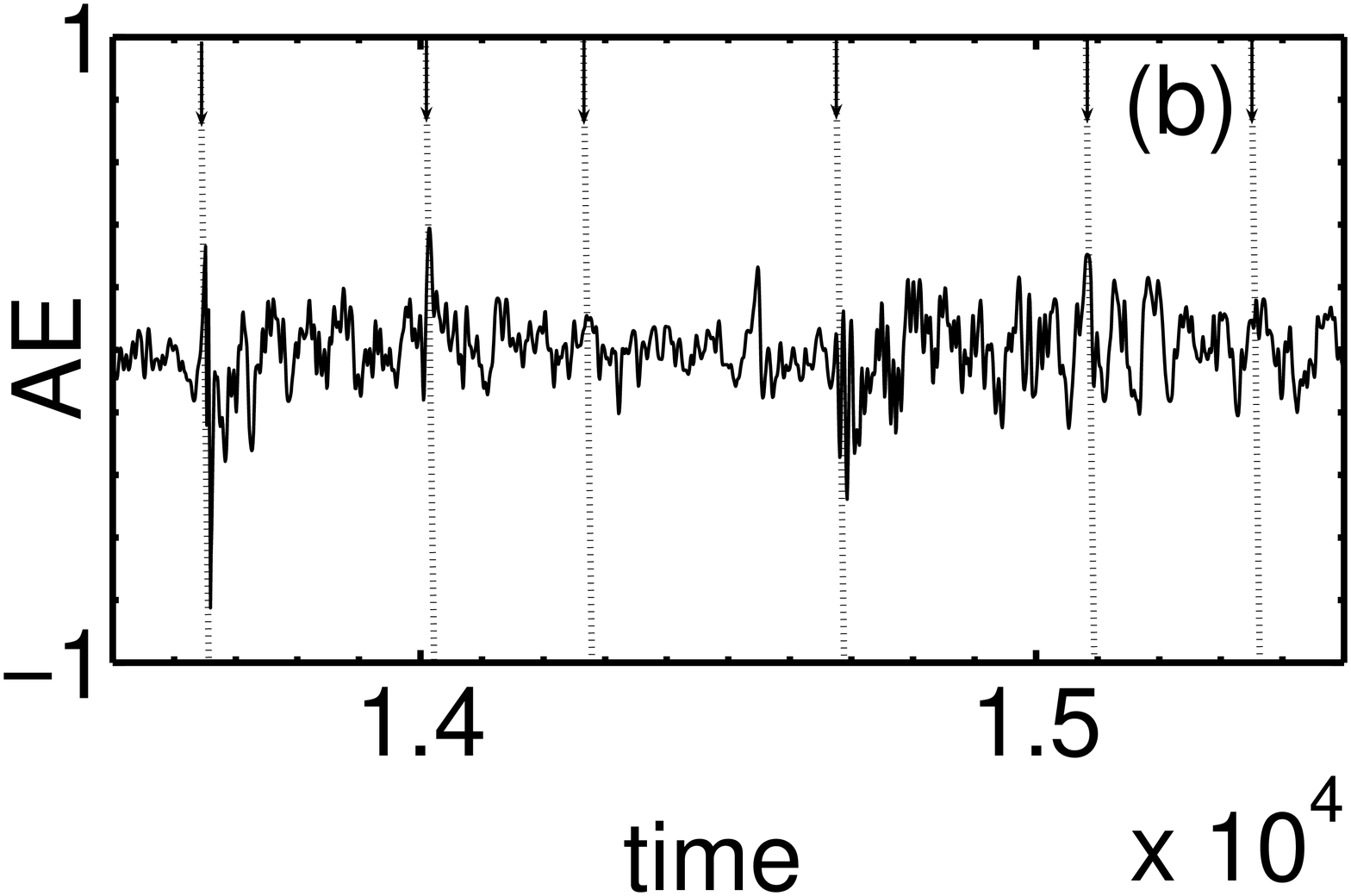}
}
\hbox{
\includegraphics[height=2.5cm,width=4.1cm]{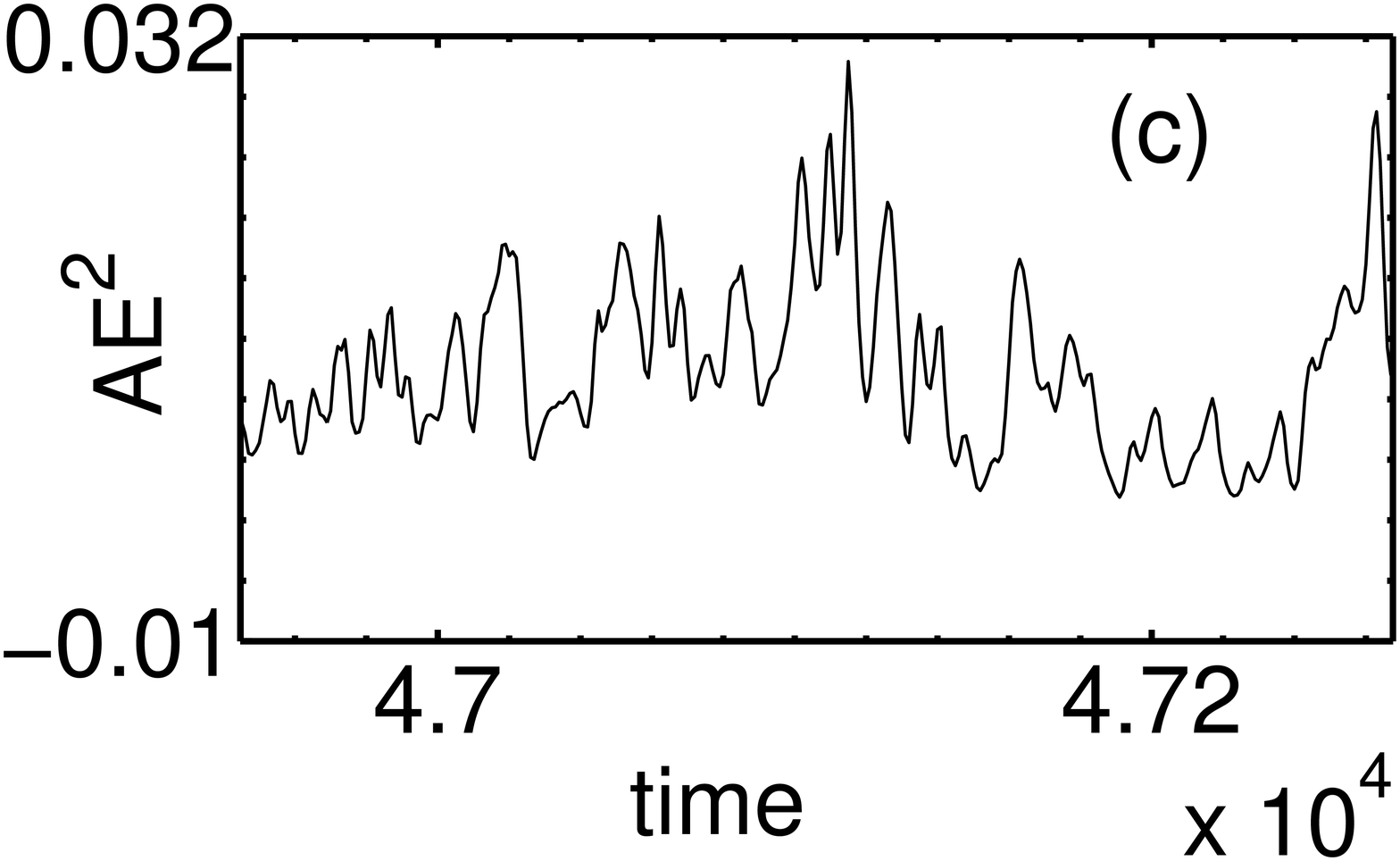}
\includegraphics[height=2.5cm,width=4.1cm]{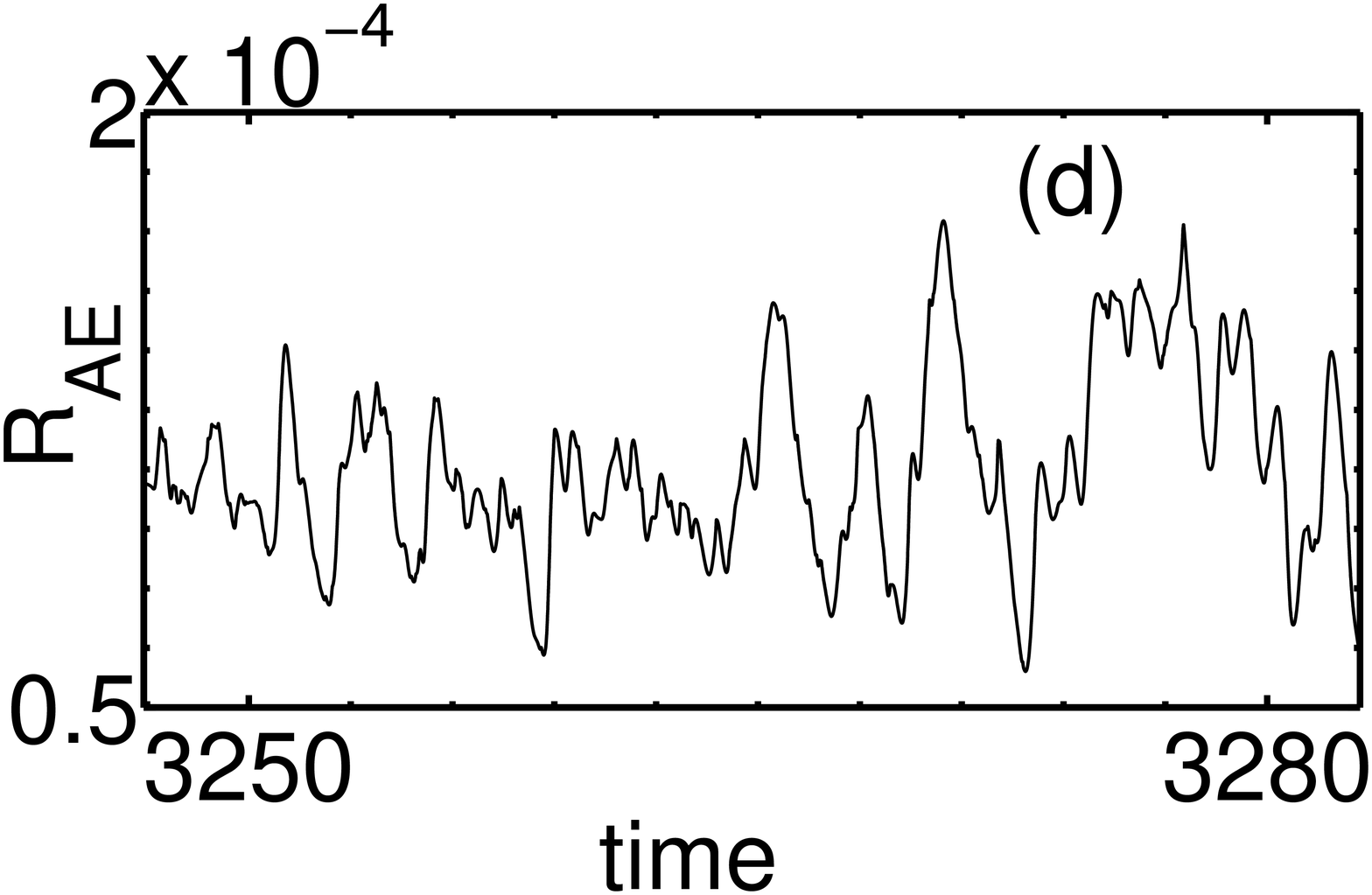}
}
}
\caption{ (a) Raw and (b) cured AE signal  for $V= 4.8 cm/s$. (c) Square of the amplitude ( in arbitrary units) for the data in (b). (d) model AE signal for $V^s = 2.48$ and $m=0.001$ which is similar to (c) except for the magnitude of fluctuations.  
}
\label{AE}
\end{figure}

To verify the prediction of chaotic dynamics, we have performed peeling experiments of an adhesive tape mounted on a roller driven at a constant traction velocity   in the wide range $0.2$ to $7.6$ cm/s.  A schematic of the experimental setup is  shown in Fig. \ref{schematic}a. An adhesive roller tape of radius $R$ is mounted on an axis passing through $O$ with a motor positioned at $O'$ that provides a constant pull speed $V$.  AE signals associated with stick-slip dynamics are  monitored using  a high quality microphone. Signals were digitized at the standard audio sampling frequency of $44.1$ kHz (having $6$ kHz band width) with $16$ bit signals stored in raw binary files.  For low pull speeds $V$,  regular AE bursts are seen that correspond to stick-slip events separated by oscillatory decaying amplitude.  With increasing pull velocity, the AE bursts become irregular and continuous as shown in Fig. \ref{AE}a. There are 38 data files each containing $1.2 \times 10^6$ points. As in most experiments on AE, signals are noisy.

Time series  analysis (TSA) begins by unfolding the dynamics through phase space reconstruction of the attractor   by embedding the time series in a higher dimensional space using  a suitable time delay\cite{GP}. 
Let $[x(k),k=1,2,3,\cdots ,N]$ be the AE signal with $\Delta t$ as the sampling time. Then, $d-$dimensional vectors are defined by   $\vec{\xi}_{k}=[x(k),x(k+\tau),\cdots,x(k+(d-1)\tau)]; \,\, k=1,\cdots ,[N-(d-1)\tau]$.
The delay time $\tau$ is either obtained from the autocorrelation function  or mutual information \cite{HKS}. Then, the chaotic nature of the attractor is quantified by establishing the existence of correlation dimension and a positive Lyapunov exponent. 

The correlation integral defined as the fraction of pairs of points $\vec{\xi}_{i}$ and $\vec{\xi}_{j}$ whose distance is less than $r$, is given by
$C(r)=\frac{1}{N_p}\sum_{i,j} \Theta(r-|\vec{\xi}_i-\vec{\xi}_j|)$,
where $\Theta(\cdots)$ is the step function and $N_p$ the number of vector pairs summed. A window is imposed to exclude temporally correlated points \cite{HKS}. If the attractor is self-similar then, $C(r)\sim r^{\nu}$, where $\nu$ is the correlation dimension \cite{GP}. Then, as $d$ is increased, one expects to find convergence of the slope $dln C(r)/d ln r$ to a finite value in the limit of small $r$. In practice,  the scaling regime is found at  intermediate length scales due to the presence of noise.

As the AE signals are noisy, we have used a modified Eckmann's algorithm suitable for noisy time series \cite{Anan97}. Briefly, Eckmann's algorithm \cite{Eckmann} relies on connecting the initial small difference vector $\vec{\xi}_i - \vec{\xi}_j$ to evolved difference vector through a set of tangent matrices. The number of neighbors used is typically min$[2d,d+4]$ contained in shell size $\epsilon_s$ defined by inner and outer radii $\epsilon_i$ and $\epsilon_0$ respectively. ( $\epsilon_i$ also acts as a noise filter.) The modification we effect is to allow more number of neighbors so that the noise statistics superposed on the signal is sampled properly. We impose  additional constraints  that the sum of the exponents be negative for a dissipative system, and also demand the existence of stable positive and zero exponents ( a necessary requirement for continuous time systems like AE) over a finite range of shell sizes $\epsilon_s$.  The algorithm works well for reasonably high levels of noise in model systems \cite{Anan97}  as also for experimental time series (for details, see Ref. \cite{Anan97}).  We have also repeated the analysis using the TISEAN package \cite{HKS}.

The data sets are first cured  using a noise reduction technique  \cite{HKS}.  Figs. \ref{AE}a and b show the raw and cured data respectively for $V= 4.8 cm/s$. Clearly, the dominant features ( the peaks shown by arrows) of the time series are retained except that the amplitude is reduced \cite{HKS}. Indeed, the two stage power law distribution for the amplitude of AE signals for the raw data are retained except that the exponent for small amplitudes is  reduced (from 0.33 to 0.24)  without altering that for the large amplitudes. The cured data are used to calculate the correlation dimension for all the data files.  However, while  raw data are adequate for calculating the Lyapunov spectrum from  our algorithm, cured  data are required  for the TISEAN package. 
To reduce the computational time, only one fifth of the total points are used. 
\begin{figure}
\hbox{
\includegraphics[height=3.1cm,width=4.7cm]{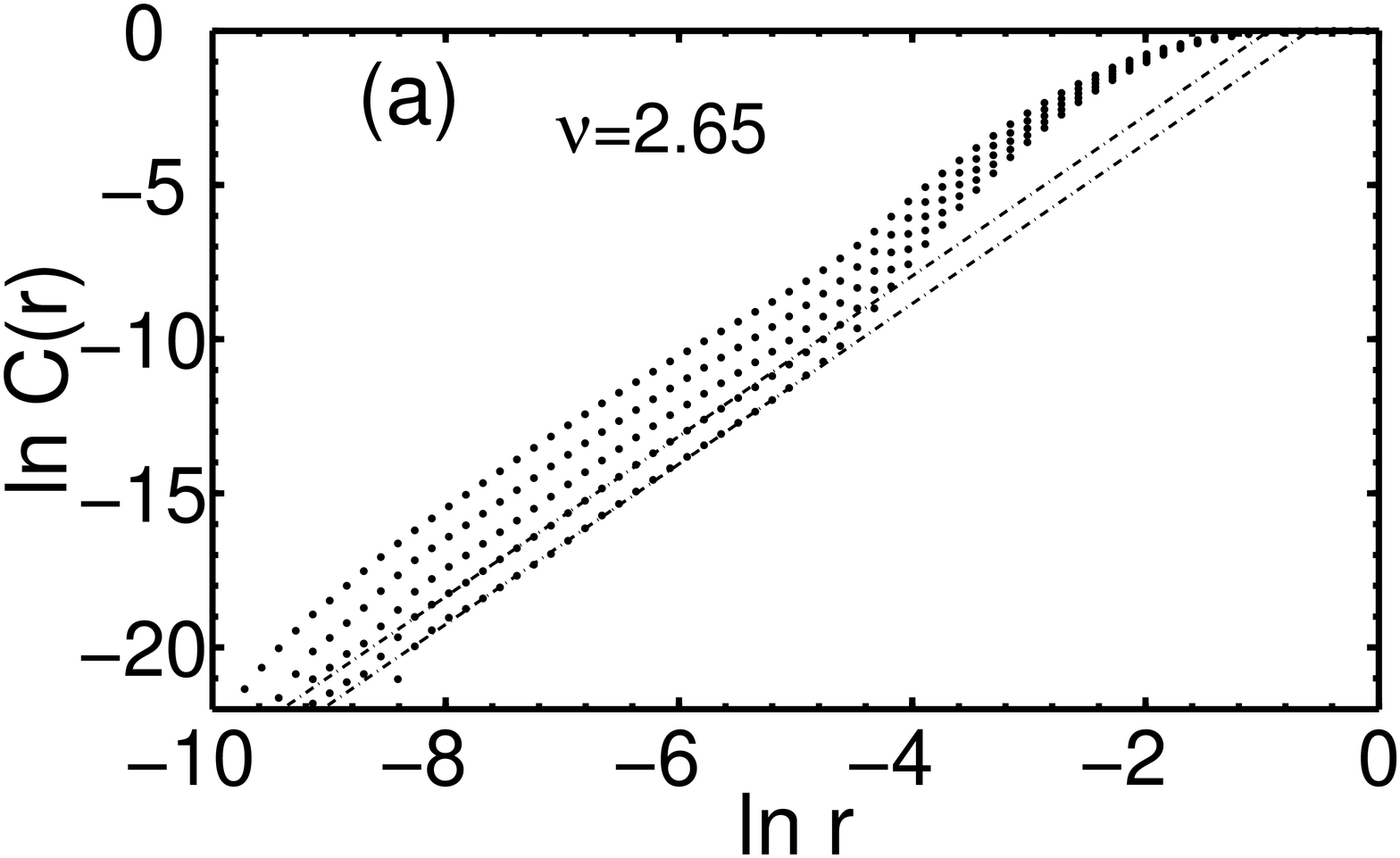}
\includegraphics[height=3.1cm,width=3.7cm]{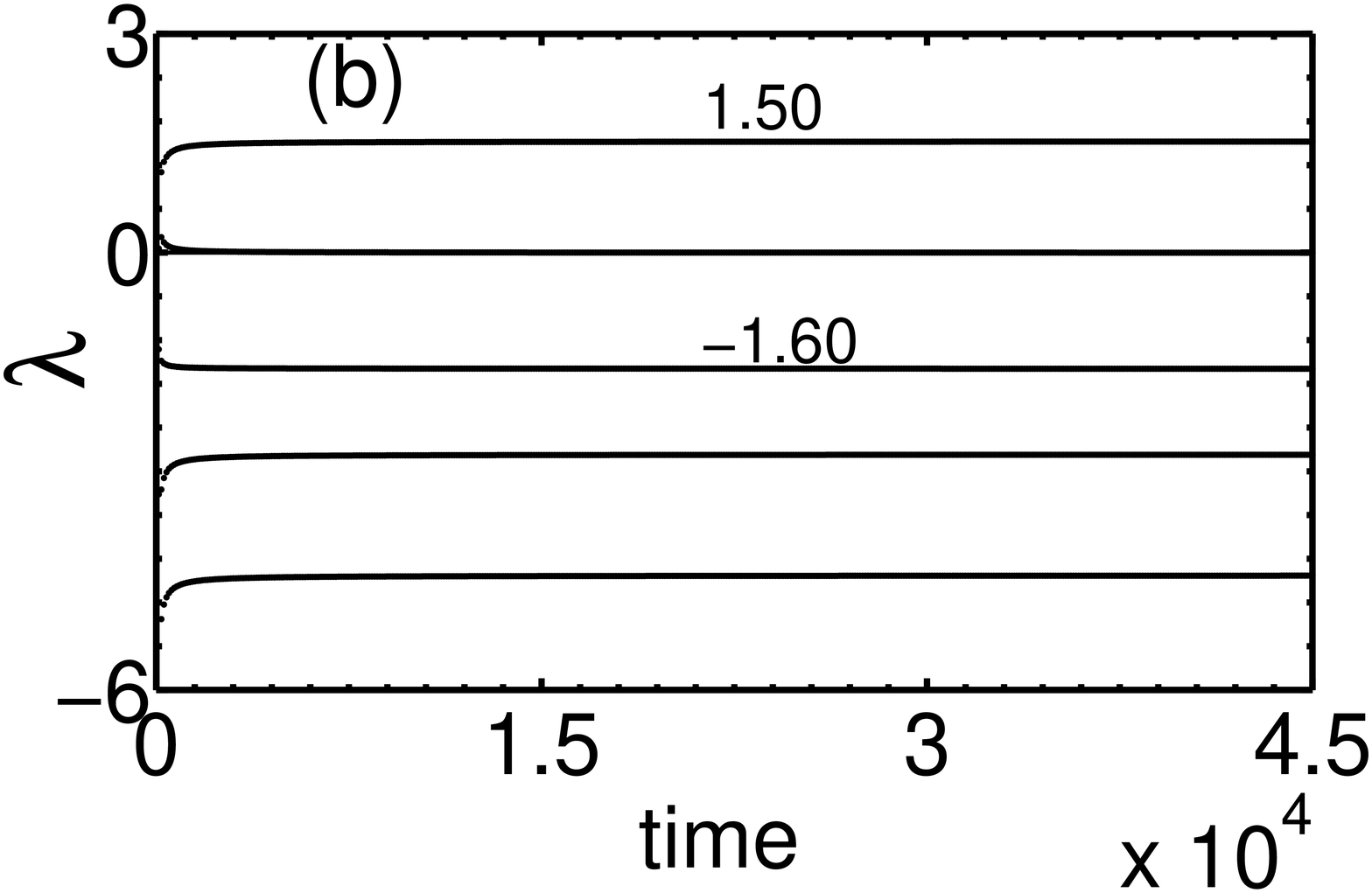}
}
\caption{ (a) Correlation integral for pull velocity $4.8 cm/s$  from $d=6$ to 10. Dashed lines are guide to eye.
(b) Lyapunov spectrum of the AE signals for traction velocities  $V= 4.8 cm/s$.}
\label{CR}

\end{figure}

The autocorrelation time is 4 units in sampling time. A smaller value of $\tau=1$ is used to calculate  $C(r)$. A log-log plot of $C(r)$  for the pull velocity $4.8 cm/s$ is shown in Fig. \ref{CR} a  for $d=6$ to $10$. A scaling regime of three orders of magnitude is seen with $\nu \sim 2.65 \pm 0.05$. However, converged values of $\nu$ ( using our method and TISEAN package) are seen, only for data sets for pull velocities from $3.8$ to $6.2cm/s$ with $\nu$ in the range $2.6$ to $2.85 \pm 0.05$. 
 
Using our algorithm, the calculated Lyapunov spectrum is shown in Fig. \ref{CR} b  for $V =4.8 cm/s$ keeping $\epsilon_o=0.065$. Note that the second exponent is close to zero as expected of continuous flow systems. We have calculated Lyapunov spectrum for the full range of traction velocities and we find (stable) positive and zero exponents \cite{error} only in the region 3.8  to 6.2cm/s, consistent with the range of converged values of $\nu$. 

As a cross-check, we have calculated  the Kaplan-Yorke dimension $D_{ky}$ from the relation $D_{ky}=j+\frac{\sum_{i=1}^j \lambda_i}{|\lambda_{j+1}|}; \sum_{i=1}^j\lambda_i>0;\sum_{i=1}^{j+1}\lambda_i<0 $. For the case shown in Fig. \ref{CR} b, we find $D_{ky}= 2+ 1.5/1.6 = 2.94$ consistent  with $\nu$ obtained from $C(r)$ \cite{error}. Similar deviations are seen for other pull velocities. The $D_{ky}$ values obtained from the TISEAN package  are uniformly closer to the $\nu$ values, typically $D_{ky} = \nu + 0.1$.  Finally, we note that the positive exponent decreases toward  the end of the chaotic domain (6.2 cm/s). These results show unambiguously that the underlying dynamics responsible for AE during peeling is chaotic in a mid range of pull speeds.

To understand the results, consider  a recent model for peeling of an adhesive tape \cite{Rumiprl}.  In  Fig. \ref{schematic}a,  the distance $OO'$ is denoted by $l$ and the peeled length of the tape $PO'$ by $L$. The angle between the tangent to the contact point $P$ ( projection of the contact line $PQ$ onto the plane of the paper) and $PO'$ is denoted by $\theta$ and the angle $\angle{POO'}$ by $\alpha$.   From Fig. \ref{schematic}a, we get $L\ {cos}\, \theta = -l\ {sin}\,\alpha$ and  $L\ {sin}\,\theta = l\ { cos}\,\alpha - R$. As the peel point $P$ moves with a local velocity $v$, the pull velocity is given by $V= v + \dot u + R \  {\rm cos}\ \theta  \ \dot \alpha$. Defining  $u(y)$ to be the displacement with respect to the uniform `stuck' peel front and defining $v(y),\theta(y)$ and $\alpha(y)$  at all points $y$ along the contact line, the above equation  generalizes to 
\begin{eqnarray}
{1\over b} \int^b_0 \big[V- v(y) -  \dot u(y)  -  R \ \ \dot\alpha(y) \ \ {\rm cos} \ \theta(y) \big]dy =0. 
\label{Vconstraint}
\end{eqnarray}
where $b$ is the width of the tape. As the contact line dynamics is controlled by the soft glue, we assume that the effective elastic constant $k_g$ along the  contact line is  much smaller than that of the tape material $k_t$. This implies that the force along $PO'$ equilibrates fast and the integrand in  Eq. (\ref{Vconstraint}) can be assumed to vanish for all $y$. 

The basic idea of the model is that while stick-slip dynamics is controlled by the peel force function $f(v)$, the associated AE is the energy dissipated during rapid movement of the peel front. We begin by defining dimensionless  variable $\tau = \omega_{u} t$, with  $\omega_{u}^2=({k_t/b \rho})$  where $\rho$ is  the mass per unit width of the length $L$.  Similarly, we define $u = X d $, $l =  l^s d$, $L =  L^s d$   and $R =  R^s d$ using a basic length scale $d=f_{max}/k_t$, where $f_{max} = f(v_{max})$ is the maximum value of the peel force function $f(v)$. We define  the scaled peel  force function by  $\phi(v^s) =f(v^s(v))/f_{max}$ (Fig. \ref{schematic}b). Here,  $v^s=v/v_c\omega_u d$ and  $V^s=V/v_c\omega_u d$ are the dimensionless peel and pull velocities  respectively with $v_c = v_{max}/ \omega_u d$. Using a scaled variable $r = y/a$, with $a$ referring to  a  unit length along the peel front, the scaled kinetic energy can be written as  $U^s_K = {1\over2 C_f}\int^{b/a}_0 \Big[\dot \alpha(r) +{v_c v^s(r)\over R^s} \Big]^2 dr + {1\over2}\int^{b/a}_0 \Big[\dot X(r) \Big]^2 dr$.  Here the first term represents the rotational kinetic energy and  the second term, the kinetic energy of stretched tape.  $C_f=(f_{max}/k_t)^2(\rho/\xi)$ represents the relative strength of the two terms, where $\xi$ is the moment of inertia per unit width of the roller tape. The potential energy  is given by $U^s_P = {1\over2}\int^{b/a}_0 X^2(r)  dr + {k_0\over2}\int^{b/a}_0 \Big[{\partial X(r) \over \partial r} \Big]^2 dr $ with $k_0 = (k_gb^2/k_ta^2)$. 
The first term arises from the displacement of the peel front due to stretching of the tape and the second term due to inhomogeneity along the front. The total dissipation is the sum of dissipation arising from the peel force function $\phi(v^s)$ and from the rapid movement of the peel front given by  ${\cal R}^s = {1\over b} \int^{b/a}_0 \int \phi(v^s(r)) dv^s dr + {1\over2}\int^{b/a}_0 \gamma_u \Big[{\partial \dot X(r) \over \partial r} \Big]^2 dr$ respectively. $\phi^s(v^s)$ is assumed to be derivable from a potential function $\Phi(v^s) = \int \phi(v^s) dv^s$.   The  second term denoted by $R_{AE}$ is the Rayleigh dissipation functional which is interpreted as the energy dissipated in the form of AE.  The scaled $\gamma_u$  is related to the unscaled dissipation coefficient $\Gamma_u$ through $\gamma_u = \Gamma_u \omega_u/(k_t a^2)$. 

The scaled local form of  Eq. \ref{Vconstraint} is
\begin{equation}
\dot X = (V^s - v^s)v_c + R^s \ {l^s \over L^s} \ ({sin}\ \alpha)\ \dot \alpha.
\label{Xdot} 
\end{equation}  
Using Lagrange equations of motion, we obtain  
\begin{eqnarray}
\label{X}
\ddot{X} &=& - X + k_0 \frac{\partial^2 X}{\partial r^2} + \frac{\phi(v^s)}{(1 + l^s/L^s \, {sin}\, \alpha)} +
\gamma_u \frac{\partial^2 {\dot X}}{\partial r^2},\\
\label{v}
v_c\dot{v}^s&=&{R^sl^s\over L^s} \{\dot \alpha^2 \big( cos \alpha -{R^s l^s}({sin \alpha \over L^s})^2 \big)  + {\ddot \alpha} sin \alpha \}-{\ddot X},\\
\label{alpha}
\ddot \alpha &=& - {v_c \dot v^s \over R^s}  - C_f R^s {l^s/L^s \, {sin}\, \alpha \over (1 + l^s/L^s \, {sin}\, \alpha)} \phi(v^s).
\end{eqnarray}

Equations  (\ref{Xdot}-\ref{alpha}) are solved using an adaptive step size stiff differential equations solver (MATLAB 'ode15s') with  open boundary conditions. The nature of the dynamics depends on the pull velocity $V^s$, the dissipation coefficient $\gamma_u$ and $C_f$. $C_f$ depends on  the roller inertia  $I =\xi b$ ($10^{-5} \le I \le 10^{-2}$) and the tape mass $m = \rho b$ ( $ 0.001\le m \le 0.1$).  $\gamma_u$ ranges from 0.001 to 0.1. Other  parameters are fixed at $R^s$=0.35, $l^s$ =3.5, $k_0=0.1$ ( $k_t =1000 N/m$) and  $N=50$. The (unscaled) peel force function $f(v)$ preserves major experimental features like the values of $f_{max}$,  $v_{max}$ and the velocity jump \cite{MB}. 

\begin{figure}
\hbox{
\includegraphics[height=3.0cm,width=4.5cm]{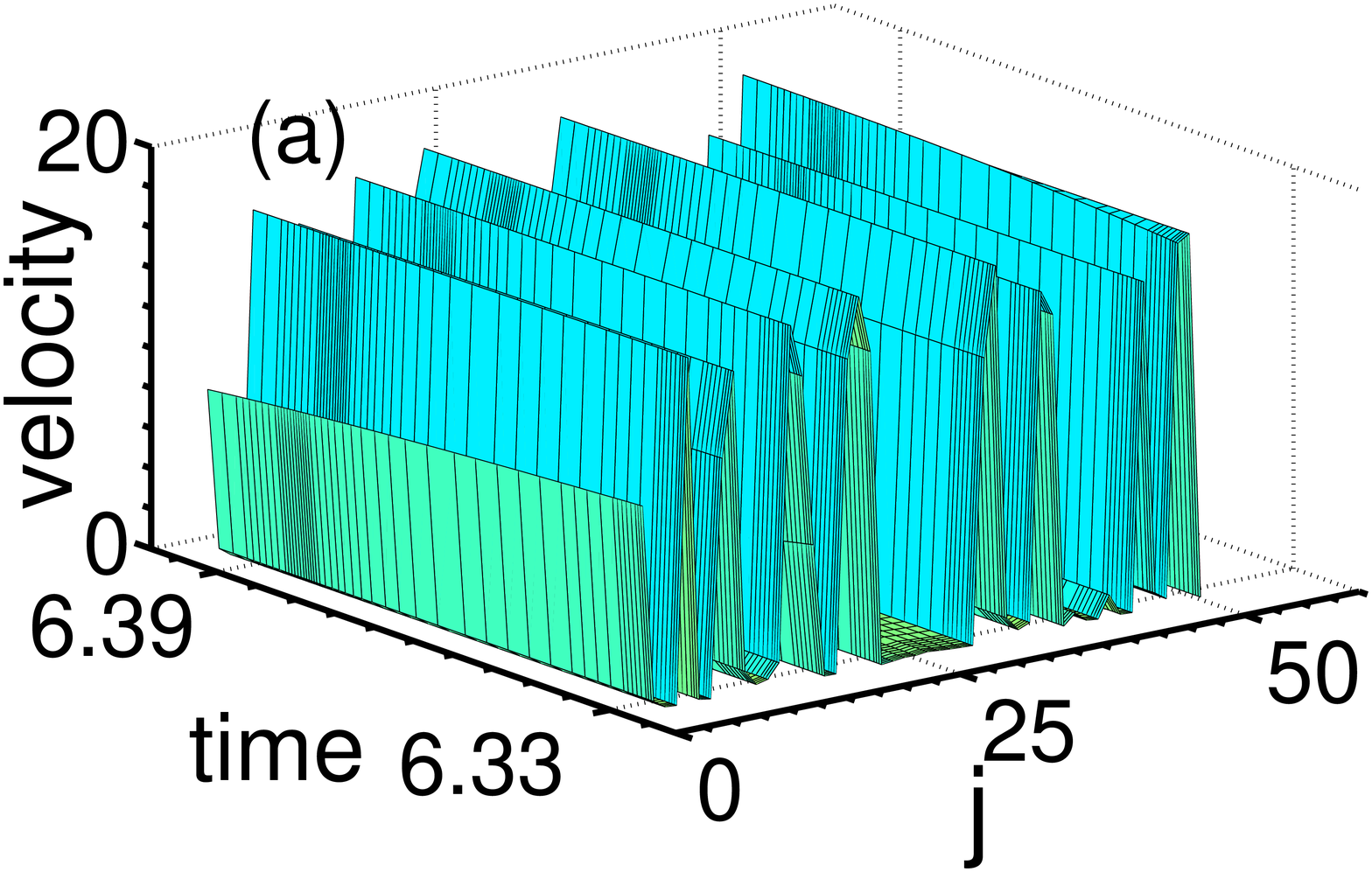}
\includegraphics[height=2.7cm,width=4.0cm]{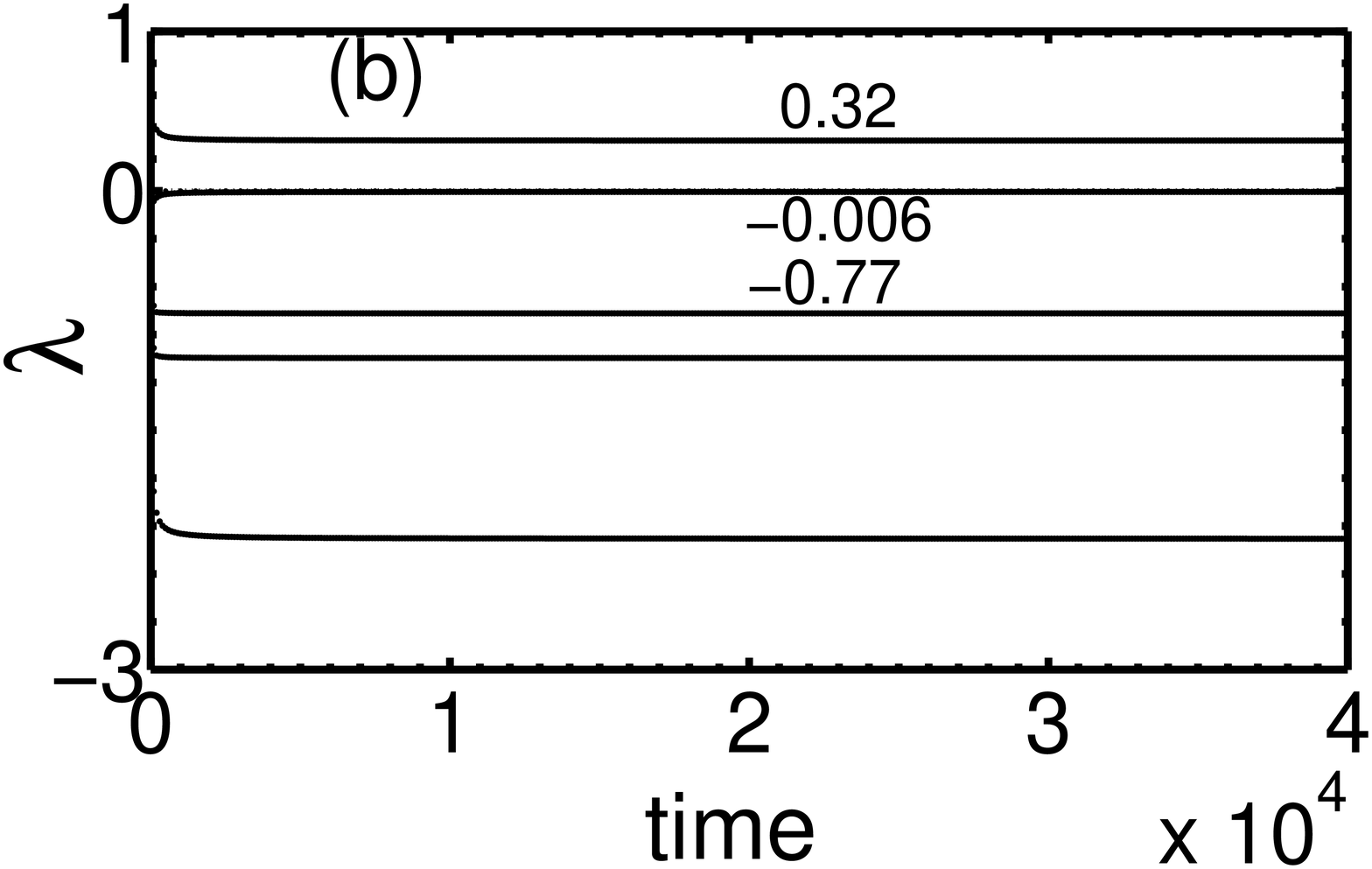}
}
\caption{(a)Stuck-peeled configuration for $V^s=2.48$ and $m=0.001$.(b)The corresponding Lyapunov spectrum for $R_{AE}$.}
\label{Lyapmodel}
\end{figure}

The results reported are for  $m=0.001$ and 0.055, $I=0.01$ and low dissipation coefficient $\gamma_u =0.01$. Physically, low $\gamma_u$, implies weak  coupling between velocities on neighboring points  on the peel front. Thus, local dynamics dominates and hence more ruggedness  leading to higher dissipation $R_{AE}$ (than for large $\gamma_u$). Indeed,  even for low $V^s$, the  peel front breaks up into  stuck and peeled segments (see Fig. \ref{Lyapmodel}a for $V^s = 2.48$ and also Ref. \cite{Rumiprl}).  Hence, the acoustic energy dissipated $R_{AE}$ is noisy.  

Several qualitative features of the experimental AE signals such as the change from burst to continuous type with pull velocity are displayed by $R_{AE}$. The observed two stage power law distribution for the experimental AE signals is reproduced by the model. For instance, for the model signal in Fig. \ref{AE}d, the exponent values are $m_E=0.6$ and 2.0 for small and large  values  respectively, consistent with the two exponents $m_A=0.24$ and 3.0 for Fig. \ref{AE}b. (Note that energy $R_{AE}$ is the square of  AE amplitude.)  Ref. \cite{Rumiprl} also reports a spatio-temporal chaotic state that corresponds to ``edge of peeling picture''  for high tape mass $m=0.1$, $I = 0.01$ and low pull speeds. However, as experimental AE signals become chaotic as a function  pull velocity (not studied in Ref. \cite{Rumiprl}),  the correct quantity to analyze is the energy dissipated in the form of AE, $R_{AE}(t)$ ( an average over the peel front).

Following the embedding technique, we have analyzed the model AE signal $R_{AE}(t)$ and computed  the correlation dimensions and Lyapunov spectrum for the entire instability domain. We find stable positive and zero exponents for a range of $\epsilon_o$ values.  A plot of the spectrum for $m=0.001$ and $V^s=2.48$ ($\epsilon_o=0.08$) is shown in Fig. \ref{Lyapmodel}b which gives  $D_{ky} = 2+ 0.32/0.77 = 2.4$ while $\nu = 2.2 \pm 0.02$ \cite{error}. Converged values of $\nu$ ranging from 2.2 to 2.7 ($D_{ky}$ in the range 2.4 to 3.0) are seen in the sub-interval $ 1.48 \le V^s\le 6.48$ of the instability along with stable positive exponents. Similar converged values of $\nu$ for $m =0.055$ ( ranging from  2.6 to 3.2 with $D_{ky}$ in the range 2.7 to 3.3) are seen in a  mid range of $V^s$. The value of the positive exponent decreases for large $V^s$.

Several conclusions emerge from the study.  First, the presence of chaos  in experimental AE signals supported by the model shows that deterministic dynamics is  responsible for AE during peeling. Second, the model also provides answers to questions raised by the TSA. For instance, the model shows that while stick-slip is controlled by the peel force function, acoustic emission $R_{AE}$ is controlled by the local kinetic energy bursts on the peel front generated during switching between the stuck and peeled states (Fig. \ref{Lyapmodel}a).  This mechanism provides insight into the transition from burst to continuous type of AE. At low pull velocities, $V^s$, the number of stuck segments are few, each containing many spatial points (Fig. \ref{Lyapmodel}a), with only a few large velocity bursts  leading to  burst type $R_{AE}(t)$. With increasing $V^s$, the number of stuck segments increases (each containing fewer points) with a large number of small local velocity bursts that  therefore lead to continuous AE signals (similar to Figs. 3b,c of Ref.\cite{Rumiprl}). Hence, the decreasing trend of the positive Lyapunov exponent with pull velocity observed in experimental  signals can be attributed to peel front breaking up into large number of small stuck-peel segments.  Thus, the model provides insight and clarifies the connection between stick-slip and the AE process. The work also addresses the general problem of extracting dynamical information from noisy AE signals.

Our study  has relevance to time dependent issues of adhesion, in particular,  to failure of adhesive joints and composites that are subject to fluctuating loads. Specifically, the analysis suggests that a larger value of the positive Lyapunov exponent (its inverse  giving the time scale) implies higher dissipation and hence earlier failure. Thus, using acoustic emission technique to monitor AE signals in these cases coupled with the estimation of the largest Lyapunov exponent  could prove to be useful. 

Many of these features are common to the  PLC effect. The effect attributed to pinning and unpinning of dislocations from solute atmosphere, is clearly, a distinct physical process from peeling. Yet, the  negative force-velocity relation and the existence of chaotic dynamics in a mid range of drive rates are seen both in experiments and a model for the PLC effect as well \cite{Anan97,Anan04,GA07}. Dynamically, the existence of chaotic dynamics as also the decreasing trend of the positive Lyapunov exponent, seen in both in the PLC effect and peeling,   is the result of a reverse forward Hopf bifurcation (HB) (end of the instability) that follows the forward HB (onset) \cite{Anan04}. As chaotic window is seen in both cases, it is likely that it  is a general feature in other stick-slip situations that are limited to a window of  drive rates.  
 
GA acknowledges the support of Grant No. 2005/37/16/BRNS.


\begin{thebibliography}{99}
\bibitem{error} Typical  error bars for the positive, zero and negative  exponents are  $\pm 0.01, \pm0.005, \pm 0.05$ respectively. $D_{ky}$ values have an error of $\pm 0.05$.
\end{thebibliography}
\end{document}